\documentclass[desactivate]{aa} 

\usepackage{graphicx}
\usepackage{subfig}
\usepackage{txfonts}
\usepackage{csvsimple,booktabs}
\usepackage{filecontents}

\usepackage{pdfpages}

\usepackage{natbib}
\usepackage[colorlinks=true,allcolors=blue]{hyperref} 
\bibpunct{(}{)}{;}{a}{}{,}

\begin{document}

  \title{Resolving core-envelope degeneracies in giant planets and sub-Neptunes: Constraining the equivalence in the presence of dilute gradients}

   \author{C. Wilkinson\inst{1}\thanks{Corresponding author: \email{christian.wilkinson@obspm.fr}} 
          \and B. Charnay\inst{1}
          \and S. Mazevet\inst{2}
          \and B. Perrier\inst{1}
          \and A.-M. Lagrange\inst{1}
          \and L. Delaye\inst{1}
          }

   \institute{
   Laboratoire d'Instrumentation et de Recherche en Astrophysique (LIRA), Observatoire de Paris, Universit\'{e} PSL, Sorbonne Universit\'{e}, Université Paris Cit\'{e}, CY Cergy Paris Universit\'{e}, CNRS, 92190 Meudon, France \\
       \and
       Observatoire de la Côte d'Azur, Universit\'{e} Côte d'Azur, 96. Boulevard de l'observatoire, 06300 Nice, France
    }

   \date{Received 10 February 2026 / Accepted 15 July 2026}

\abstract
    {The widespread recognition of dilute, `fuzzy' cores in giant planets and massive volatile mantles in sub-Neptunes has transformed planetary interior modelling into a highly degenerate, multi-dimensional problem.}
    {To map the structural degeneracies introduced by these complex interiors, we present \texttt{fuzzycore}, an open-source static structural integrator with a unified layered architecture spanning rock-water-envelope sub-Neptunes through volatile-rich gas giants. The framework supports dense parameter sweeps over compositional gradients at fixed observed $(M_\mathrm{p}, R_\mathrm{p}, Z_\mathrm{atm})$, enabling consistent forward-model exploration of interior degeneracies across the radius valley and the warm-giant population.}
    {The framework solves the equations of hydrostatic equilibrium across phase-separated iron, rock, and water layers beneath a gaseous envelope. We employed a highly resolved adaptive grid to model arbitrary heavy-element gradients within the envelope, parameterising the dilution width to systematically evaluate the structural impact of varying core-envelope boundaries.}
    {We benchmarked \texttt{fuzzycore} against an evolution-derived Jupiter profile, demonstrating that when envelope boundary metallicities and integrated heavy-element budgets are matched, the macroscopic planetary radius is robust to the exact functional topology of the gradient to within 2.3\%. This empirical robustness allowed us to use smooth parameterisations to densely map interior degeneracies. Applying this to the sub-Neptune regime, we generated a water-world degeneracy atlas, quantifying how the required envelope metallicity and dilution width trade off against assumed water-mass fractions to reproduce a fixed planetary radius.}
    {By decoupling structural profiling from time-dependent energy transport and convective mixing, \texttt{fuzzycore} enables the systematic mapping of static degeneracies across the sub-Neptune and warm-giant populations at resolutions that complement, rather than replace, fully evolutionary Henyey treatments. It serves as a forward-model backend for linking atmospheric metallicity priors from JWST and \textit{Ariel} to deep interior architectures.}


    \titlerunning{Resolving core-envelope degeneracies in giant planets and sub-Neptunes}
   \authorrunning{Wilkinson et al.}

   \maketitle
\ifdefined\nolinenumbers\nolinenumbers\fi

\section{Introduction}
\label{sec:intro}

The internal structure of giant and sub-giant planets remains one of the most significant sources of degeneracy in planetary science. A central challenge of modern exoplanetary astrophysics is interpreting the bimodal distribution of small planetary radii often referred to as the `radius valley' \citep{fulton_california-_2017}. Bridging this valley requires determining whether planets occupying the sub-Neptune regime possess thick hydrogen-helium (H/He) atmospheres, massive supercritical water oceans, or some combination thereof \citep{dorn_hidden_2021}. Consequently, the specific spatial distribution of heavy elements within the interior significantly affects the total radius and gravity moments ($J_n$) of the planet. 

Planetary evolution theory, bolstered by high-precision gravity data from \textit{Juno} and \textit{Cassini} has shifted away from distinct solid surface models towards a `fuzzy core' approach \citep{stevenson_interiors_1982, bolton_junos_2017, wahl_comparing_2017}. Wherein planetary cores exist not as sharp physical boundaries, but rather as dilute mixtures of rock, water, and hydrogen-helium fluid \citep{helled_internal_2018, brouwers_how_2018, helled_revelations_2022}. To capture these complex interiors, evolutionary frameworks utilising the Henyey method, a fully implicit relaxation scheme that discretises the equations of stellar structure and solves them simultaneously on the whole grid through Newton--Raphson iteration \citep{henyey_new_1964}, have been successfully adapted to planetary applications, notably beginning with \citet{vazan_convection_2015} and the inclusion of mixing in \texttt{MESA} \citep{muller_challenge_2020}. Contemporary fully implicit codes such as \texttt{APPLE} \citep{sur_apple_2024} build upon this foundation to track the time-dependent mixing and cooling of these inhomogeneous structures over billions of years.

Furthermore, internal structure serves as a fossil record of a planet's formation history. The two dominant giant planet formation paradigms, core accretion (CA) and gravitational instability (GI), are expected to yield distinct internal architectures. Standard CA involves the slow buildup of a solid core followed by rapid gas accretion \citep{pollack_formation_1996}, naturally lending itself to metal-enriched deep interiors. As accreting planetesimals and pebbles dissolve in the envelope, they naturally form the extended compositional gradients and fuzzy cores our framework models \citep{lozovsky_jupiters_2017, brouwers_how_2018}. In contrast, GI forms planets via the rapid fragmentation of the protoplanetary disc \citep{boss_giant_1997}, theoretically producing more homogeneously mixed bodies of solar composition, though late-stage planetesimal bombardment can introduce subsequent heavy-element enrichment \citep{helled_core_2008}. Accurately mapping the internal heavy-element distribution is therefore crucial for breaking these structural degeneracies and disentangling competing formation pathways.

Simultaneously, the proliferation of low-density sub-Neptunes and `super-puff' discoveries, contrasted against planets with massive supercritical oceans, requires numerical methods that can bridge the gap between low-pressure extended atmospheres and extreme-pressure condensed phases without failing at numerical discontinuities \citep{bean_nature_2021, luque_density_2022}. In this work, we present \texttt{fuzzycore}\footnote{The open-source Python framework is publicly available at \url{https://github.com/ChristianSWilkinson/fuzzycore}}, a static, multi-zone hydrostatic solver designed for the fast forward-model exploration of static interior degeneracies. \texttt{fuzzycore} is not an evolutionary code and does not model time-dependent mixing, self-consistent energy transport, or phase-equilibrium critical curves; for these questions, we defer to established Henyey-class frameworks (\texttt{MESA}, \texttt{APPLE}, \texttt{CEPAM}, etc.) and to ab initio miscibility studies. Its niche is to densely sample the space of static structures consistent with given observations, including the dilution-width dimension $\sigma$ that evolutionary codes cannot afford to sweep at high resolution, and to explicitly map the resulting degeneracies across both the gas-giant and sub-Neptune regimes.

\section{Methods and physical model}
\label{Sec:Methods and Physical Model}

\subsection{Multi-zone hydrostatic equilibrium}
The framework determines the interior structure by solving the standard system of 1D spherically symmetric equations for mass conservation and hydrostatic equilibrium \citep[e.g.][]{guillot_giant_2015}:
\begin{equation}
\frac{dm}{dr} = 4\pi r^2 \rho(r),
\end{equation}
\begin{equation}
\frac{dP}{dr} = -\frac{Gm(r)}{r^2} \rho(r),
\end{equation}
where $m(r)$ is the enclosed mass, $\rho(r)$ is the local mass density, $P(r)$ is the pressure, and $G$ is the gravitational constant. The interior is radially differentiated into up to three distinct structural zones depending on the targeted planetary architecture: an inner refractory core, an intermediate condensed volatile mantle, and a fluid envelope. To accurately resolve the steep density gradients across material interfaces without incurring prohibitive computational costs, the spatial step size $dr$ is dynamically adapted during the integration based on the local pressure scale height and density contrast.

\subsection{Mixed iron-rock core formulation}
Standard multi-layer models often approximate the deep interior as a pure silicate core. However, at extreme pressures ($P > 10$\,Mbar) typical of giant planet centres, high-pressure water phases (such as the symmetric, hydrogen-bonded Ice X phase or superionic configurations) can exceed the density of pure, uncompressed silicates \citep[e.g.][]{mazevet_melting_2015, mazevet_ab_2019}.

To correct this and ensure strict monotonic density stratification, we implemented a mixed-core equation of state (EOS) by employing the additive volume law (Amagat's law of ideal mixing). This universal mixing rule assumes that the specific volumes of individual components combine linearly at a given pressure and temperature. We established this principle here for the solid core and subsequently applied it to all fluid mixtures throughout the framework:
\begin{equation}
    \frac{1}{\rho_{\text{mix}}(P, T)} = \sum_i \frac{X_i}{\rho_i(P, T)},
\label{eq:mixing}
\end{equation}
where $X_i$ is the prescribed mass fraction of each component. 

For the deep refractory core, we employed state-of-the-art ab initio molecular dynamics tables. We mixed an ab initio equation of state for iron \citep{bouchet_ab_2013} with a rigorous multi-phase rock formulation. Because post-perovskite $\mathrm{MgSiO_3}$ is known to dissociate at the multi-megabar pressures characteristic of giant planet cores, we explicitly separated the silicate phase into its constituent components. We utilised dedicated ab initio tables to capture the complex high-pressure melting and metallisation behaviours of $\mathrm{MgSiO_3}$ \citep{mazevet_melting_2015} and $\mathrm{MgO}$ \citep{musella_physical_2019}. The overall iron mass fraction is dictated by $X_{\text{Fe}}$ (e.g. $X_{\text{Fe}} = 0.33$ for an Earth-like refractory composition; \citealt{adibekyan_compositional_2021}).

Rather than assuming a potentially unphysical isothermal profile for this deep interior, the framework integrates the solid core using a parametrised adiabatic temperature gradient ($\nabla_{\text{ad}}$). While the framework thermally links the core to the overlying envelope, it deliberately avoids computationally prohibitive thermal micro-stepping across the phase boundary, as minor residual temperature discontinuities have a negligible impact on the integrated solid mass. By utilising these density functional theory and molecular dynamics tables, the framework accurately captures the complex non-ideal interactions and density profiles of the deep core, ensuring thermodynamic rigour and physical consistency at the extreme pressures mapping the core-envelope boundary.

\subsection{Fluid envelope mixture formulation}
To model the gaseous envelope and its dilute heavy-element gradients, we constructed a blended fluid equation of state. Following the additive volume law defined in Eq.~(\ref{eq:mixing}), the density of the ternary mixture (hydrogen, helium, and heavy elements) is calculated through ideal linear mixing of their specific volumes. Specific entropy is similarly mixed linearly. 

While the dissociation of $\mathrm{MgSiO_3}$ at multi-megabar pressures yields $\mathrm{MgO}$ and $\mathrm{SiO_2}$ that could theoretically mix into the fluid, we modelled the leaked heavy-element fraction ($Z$) in the envelope using a wide-range $\mathrm{H_2O}$ equation of state proxy. Although analytical effective-volume formulations for $\mathrm{SiO_2}$ diluted in H/He exist \citep{soubiran_properties_2016}, they are strictly valid only up to $15\,000$\,K. Because our integration sweeps explore deep interior conditions reaching up to $100\,000$\,K, a full-range EOS capable of capturing the transition to a fully ionised plasma is required to maintain thermodynamic stability and physical realism in the deep envelope. 

Utilising $\mathrm{H_2O}$ as this proxy is further justified because oxygen is the most abundant heavy element in the protoplanetary disc, making water the dominant condensable mass accreted by giant planets forming beyond the snow line \citep{mousis_irradiated_2020}. It acts as an excellent structural baseline between dense refractory silicates and lighter volatile species, accurately capturing the mean molecular weight ($\mu$) penalty and bulk compressibility without requiring computationally prohibitive, restricted-domain multi-component fluid mixtures \citep{baraffe_structure_2008, thorngren_mass-metallicity_2016}.

For local mass fractions $X$ (hydrogen), $Y$ (helium), and $Z$ (heavy elements), subject to $X + Y + Z = 1$, the hydrogen and helium fractions are dynamically scaled to preserve a prescribed baseline helium-to-hydrogen mass ratio:
\begin{equation}
    X = (1 - Y_{\text{ratio}})(1 - Z), \quad Y = Y_{\text{ratio}}(1 - Z),
\end{equation}
where $Y_{\text{ratio}} = Y / (X + Y)$ is typically set to a baseline solar value of $0.26$ \citep{asplund_chemical_2009}. By generating 2D interpolators over this blended phase space, the framework seamlessly evaluates the thermodynamic state across the entire envelope, rigorously capturing the $\mu$-compaction effects as the fluid transitions from a pristine H/He atmosphere to a highly enriched basal layer.

\subsection{Condensed volatile mantle formulation}
\label{sec:mantle}
For planets occupying the volatile-rich sub-Neptune regime (e.g. `water worlds'), the framework supports the inclusion of an intermediate condensed mantle layer. This layer represents a massive, supercritical, or high-pressure solid water phase (such as Ice VII, Ice X, or superionic water) situated between the refractory core and the gaseous envelope.

Unlike the dilute heavy elements suspended in the upper atmosphere, the mantle is modelled as a distinct, phase-separated bulk layer using a pure water equation of state \citep{mazevet_ab_2019}, incorporating the revised non-ideal entropy of the updated arXiv version (v2) of that work. The thermodynamic state of this mantle is strictly coupled to the base of the overlying envelope. The solver enforces continuity of pressure and temperature across this boundary, integrating the pure water adiabat downwards from the envelope-mantle interface ($P_{\text{int}}$) to the core-mantle boundary ($P_{\text{cmb}}$).

The architecture adopted here, in which a phase-separated condensed water mantle underlies a compositionally graded H/He--water envelope, is physically motivated by the location of the water--H/He demixing critical curve at sub-Neptune interior conditions. Ab initio molecular dynamics calculations \citep{soubiran_miscibility_2015} indicate that water and H/He form a single miscible fluid phase above their critical curve, supporting the diffuse $Z(x)$ gradient we modelled in the envelope (Sect.~\ref{sec:methods_gradients}). Below this critical curve, water exsolves into a distinct condensed phase, which we represent as the bulk water mantle. The interface pressure $P_{\mathrm{int}}$ between the two layers should therefore physically correspond to crossing the demixing curve at the local envelope temperature, and the basal envelope metallicity $Z_{\mathrm{core}}$ should be bounded above by the local water saturation in H/He at those conditions.

We emphasise that \texttt{fuzzycore} does not evaluate the water, H/He demixing critical curve self-consistently. Both $P_{\mathrm{int}}$ (returned by the shooting algorithm of Sect.~\ref{sec:nested}) and $Z_{\mathrm{core}}$ are user-controlled inputs, and we do not verify a posteriori that the converged $(P_{\mathrm{int}}, T_{\mathrm{int}}, Z_{\mathrm{core}})$ triple sits below the demixing curve in the miscible regime. The structural atlases generated in this work should therefore be interpreted as maps of mechanically permitted hydrostatic configurations; the further question of whether a given parameter triple is thermodynamically self-consistent with the demixing physics is deferred to ab initio phase-equilibrium studies and to the evolutionary codes that incorporate them. However physically correct models could be extracted from the aforementioned map.

\subsection{Compositional gradients and sub-grid conservation}
\label{sec:methods_gradients}
Rather than assuming a discrete core-envelope boundary, \texttt{fuzzycore} models the heavy element mass fraction $Z$ using a continuous compositional gradient. We adopted a Gaussian decay function as a tractable, single-parameter representation of the heavy-element distribution in the envelope:
\begin{equation}
Z(x) = Z_{\text{base}} + (Z_{\text{core}} - Z_{\text{base}}) \exp\left( -\frac{(x - 1)^2}{2\sigma^2} \right),
\end{equation}
where $x$ is the normalised mass coordinate ($x=1$ at the core boundary, $x=0$ at the surface), $Z_{\text{base}}$ is the atmospheric metallicity, $Z_{\text{core}}$ is the metallicity at the deep boundary, and $\sigma$ controls the physical extent of the dilution.

We make no claim that this smooth functional form derives directly from first-principles diffusion physics. Instead, we treated the Gaussian shape as a convenient, bounded mathematical parameterisation whose validity is established empirically in Sect.~\ref{sec:robustness} (Functional-form robustness), where we demonstrate that macroscopic planetary radii are highly degenerate with respect to the exact functional topology of the gradient.

To ensure numerical stability as the compositional transition becomes sharp ($\sigma \to 0$), the framework employs an analytical sub-grid averaging technique. We defined the grid resolution in the normalised mass coordinate as $\Delta x = 1/(N-1)$, where $N$ is the number of discrete numerical layers. When the width of the compositional gradient $\sigma$ is smaller than the grid resolution, the maximum metallicity at the inner boundary ($Z_{\text{core}}$) is dynamically downscaled to a conserved average value, $Z_{\text{dynamic}}$, to prevent unphysical density spikes:
\begin{equation}
Z_{\text{dynamic}} = Z_{\text{base}} + (Z_{\text{core}} - Z_{\text{base}}) \min\left(1, \frac{\sigma\sqrt{\pi/2}}{\Delta x}\right).
\end{equation}
Using this conserved amplitude, the heavy element distribution is evaluated over the numerical layers. As $\sigma \to 0$, $Z_{\text{dynamic}}$ smoothly approaches $Z_{\text{base}}$, transitioning the profile towards a well-mixed adiabatic state without triggering the massive pressure discontinuities typically associated with undersampled gradients.

\subsection{Compositional staircase and the stepper method}
While the Gaussian function analytically prescribes the global shape of $Z(x)$, integrating this continuous profile through the fluid equations of state requires numerical discretisation. To achieve this, the envelope is segmented into a discrete `staircase' profile consisting of $N$ layers. Within each layer $i$, the mass fraction $Z_i$ is held constant, representing a fully mixed convective zone.

To integrate the thermodynamic state through these layers, we utilised a robust adiabat stepper method. Rather than relying on simple 1D interpolations that often fail near phase transitions, the stepper operates via local linear least-squares regressions over a KD-Tree \citep{bentley_multidimensional_1975, virtanen_scipy_2020} of the 3D phase space $(P, T, S)$. For a target pressure $P_{i+1}$ and a constant layer specific entropy $S_i$, the stepper solves for the resultant temperature $T_{i+1}$:
\begin{equation}
T_{i+1} = T_{\text{mean}} + \frac{S_i - S_{\text{mean}} - \left(\frac{\partial S}{\partial P}\right)_T (P_{i+1} - P_{\text{mean}})}{\left(\frac{\partial S}{\partial T}\right)_P}.
\end{equation}
This regression-based stepping allows the solver to advance fluid adiabats accurately even in sparse or highly non-linear regions of the EOS grids.

This numerical staircase parameterisation is physically motivated by, but does not claim to quantitatively reproduce, the layered semi-convection regimes identified in hydrodynamical simulations \citep{leconte_new_2012, wood_new_2013}. A self-consistent hydrodynamic treatment of layer formation, merger timescales, and the survival of stratification against rotational and convective erosion is strictly the domain of evolutionary modelling \citep{vazan_evolution_2016, sur_apple_2024, knierim_convective_2024}. Consequently, we treated the staircase here solely as a parametric stand-in. The ultimate effect of this discretisation on the converged macroscopic radius is formally bounded by our robustness analysis in Sect.~\ref{sec:robustness}.

\subsection{Adaptive atmospheric downsampling}
When evaluating sub-Neptune planets with massive volatile mantles, the gaseous envelope can become barometrically thin ($P_{\text{int}} / P_{\text{surf}} \lesssim 10$). Forcing a high-resolution $N$-layer compositional staircase into a narrow pressure domain forces the integrator to take minuscule $\Delta P$ steps. In this regime, the numerical noise of the 2D EOS interpolation can artificially dominate the physical signal, occasionally causing chaotic temperature divergence.

To prevent this numerical instability, we implemented an adaptive downsampling algorithm. If the total logarithmic pressure span of the envelope falls below a critical threshold, the framework dynamically discards the high-resolution gradient and collapses the envelope into a heavily reduced set of composition layers. This automatically restores the stability of the root-finding algorithms by ensuring that $\Delta P$ steps remain sufficiently large to yield meaningful, monotonically stable thermodynamic gradients across the shallow envelope.

\subsection{The nested boundary-shooting method}
\label{sec:nested}
For planets modelled with distinct volatile boundaries, the interior is constrained by the atmospheric boundary conditions ($P_{\text{surf}}, T_{\text{surf}}$), creating a rigorous dependency between the envelope's specific entropy, the mantle's thermodynamic state, and the deep core's temperature. We resolved this boundary value problem using a nested shooting method.

For a given central pressure guess $P_\mathrm{c}$:
\begin{enumerate}
\item Guess interface pressure: The solver postulates a pressure $P_{\text{int}}$ for the boundary between the gaseous envelope and the dense interior.
\item Envelope integration (top-down): The envelope is integrated downwards from $P_{\text{surf}}$ to $P_{\text{int}}$, yielding the boundary temperature $T_{\text{int}}$ and entropy.
\item Entropy matching: The mantle specific entropy $S_{\text{mantle}}$ is strictly anchored to these interface conditions.
\item Core temperature projection: A pilot integration of the refractory core estimates the core-mantle boundary pressure ($P_{\text{cmb}}$). The mantle adiabat is then projected downwards to this pressure to evaluate the boundary temperature: $T_{\text{core}} = T(P_{\text{cmb}}, S_{\text{mantle}})$.
\item Core integration (bottom-up): the iron-rock core is integrated outwards using $T_{\text{core}}$ until $P_{\text{int}}$ is reached.
\item Root finding: A continuous root-finding algorithm (Brent's method) refines $P_{\text{int}}$ iteratively until the mass and pressure continuity constraints are matched perfectly.
\end{enumerate}

Crucially, the root-finding architecture is equipped with dual convergence modes. While standard interior integrations iterate the central pressure ($P_\mathrm{c}$) to match a known total planetary mass ($M_\mathrm{p}$), the framework can alternatively solve for a target surface gravity ($g$). In this gravity-targeting mode, the solver dynamically adjusts the deep interior architecture until the integrated mass and radius yield the exact gravitational acceleration required at the atmospheric boundary, enabling self-consistent linkage to standalone atmospheric radiative transfer models.

\subsection{Phase transitions and the isothermal guard}
A critical vulnerability in static interior modelling occurs at low-pressure volatile phase transitions. Ab initio water tables, such as the \citet{mazevet_ab_2019} multi-phase EOS, often struggle to smoothly resolve the transition from a high-entropy vapour to a low-entropy liquid. If a numerical step lands precisely on a jagged phase boundary, the local interpolator can yield an unphysical negative derivative $\left( \frac{\partial T}{\partial P} \right)_S < 0$, causing the model to erroneously cool the planet as pressure increases.

We addressed this with an isothermal guard. During integration, we enforced strict thermodynamic monotonicity. If the predicted subsequent temperature $T_{i+1}$ drops significantly below $T_i$, we invoked an assumption of extreme localised mixing and clamped the layer: $T_{i+1} = \max(T_{i+1}, T_i)$. The EOS density is then re-queried at this stabilised isothermal state. This allows the integrator to safely bridge volatile phase transitions without halting, while preserving the accurate adiabatic temperature gradients provided by the corrected EOS tables in the stable fluid regimes.

\subsection{Scope and limitations of the static framework}
\label{sec:methods_limitations}
Because \texttt{fuzzycore} is designed explicitly for the rapid mapping of static structural degeneracies, it possesses several fundamental limitations compared to Henyey-class evolutionary codes. First, the framework does not compute time evolution. It cannot model cooling tracks, thermal contraction histories, or the time-dependent erosion of primordial composition gradients. Second, it does not self-consistently solve the equations of energy transport to determine the boundaries of convective, radiative, or semi-convective zones based on local opacities; the internal temperature gradient $\nabla$ is prescribed parametrically. Third, it does not simulate the dynamic hydrodynamic mixing of heavy elements. Fourth, the framework does not evaluate phase-equilibrium critical curves. The location of the water--H/He demixing boundary, which physically motivates the separation between the condensed water mantle and the diffuse $Z(x)$ envelope in Sect.~\ref{sec:mantle}, is not computed self-consistently; the interface pressure $P_{\mathrm{int}}$ and the basal envelope metallicity $Z_{\mathrm{core}}$ are user-controlled and should be checked against ab initio miscibility calculations \citep[e.g.][]{soubiran_properties_2016} for physical consistency. Configurations in which $Z_{\mathrm{core}}$ exceeds the local water saturation in H/He, or in which $P_{\mathrm{int}}$ sits well inside the miscible regime, are mechanically permitted by the static solver but thermodynamically inconsistent. Consequently, while \texttt{fuzzycore} excels at densely evaluating whether a specific core-envelope architecture is statically permitted by a given planetary mass and radius, questions regarding whether such an architecture could realistically form, remain stable against convective mixing, or survive over billions of years must be directed towards complementary evolutionary models.

\section{Validation and benchmarks}
\label{Sec:VandB}

\subsection{Static benchmarking and functional-form checks}
\label{sec:robustness}

Two complementary tests established the static framework's quantitative
accuracy and its insensitivity to the assumed $Z(x)$ functional form.

\paragraph{Static Vazan benchmark.}
To anchor \texttt{fuzzycore}'s output against an established evolutionary
solver, we converted the 4.55\,Gyr Jupiter compositional profile published by \citet{vazan_jupiters_2018} from their reported $Z(r/R)$ into our normalised mass coordinate $Z(x)$, and supplied this profile directly to the static solver with $M_\mathrm{p} = 317.8\,M_\oplus$, $M_\mathrm{core} = 0.1\,M_\oplus$, $T_\mathrm{surf} = 165$\,K, and matched envelope boundary metallicities $Z_\mathrm{base} = 0.0816$, $Z_\mathrm{core} = 0.240$. The converged radius is $9.87\,R_\oplus$ against the observed Jupiter equatorial radius of $11.21\,R_\oplus$, a residual gap of approximately $12\%$. This offset is attributable to documented differences in the underlying equations of state: \citet{vazan_jupiters_2018} use \citet{saumon_equation_1995} (known as SCVH) for H/He, while \texttt{fuzzycore} uses \citet{chabrier_new_2019}, and the water proxy adopted here is that of \citet{mazevet_ab_2019} rather than the heavy-element EOS used in that work. \citet{vazan_jupiters_2018} themselves note that SCVH is `incompatible with more recent EOS calculations' (their Sect.~3.4); the static benchmark therefore validates the structural integrator without conflating it with EOS choice.

\begin{figure*}[ht!]
\centering
\includegraphics[width=0.95\textwidth]{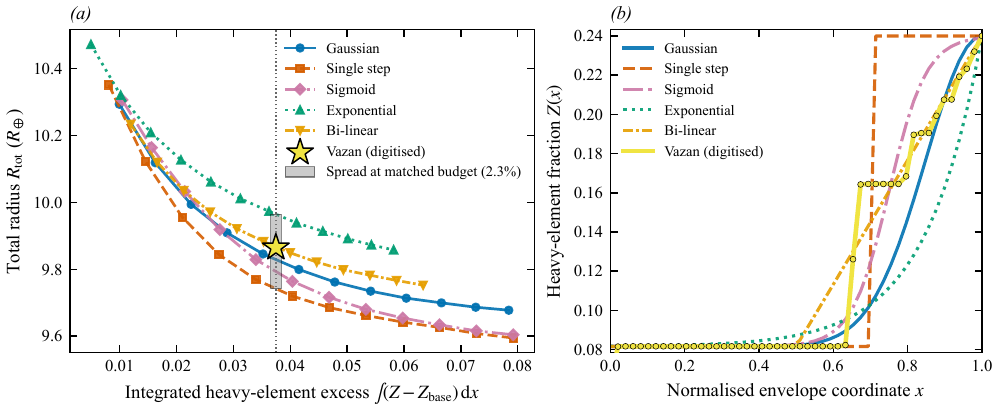}
\caption{Functional-form robustness test for the Jupiter benchmark.
\textit{(a)} Converged total radius against integrated heavy-element excess for six $Z(x)$ parametric families plus the digitised \citet{vazan_jupiters_2018} staircase, all sharing matched envelope boundary metallicities ($Z_\mathrm{base} = 0.0816$, $Z_\mathrm{core} = 0.240$) and identical bulk inputs. The shaded rectangle marks the spread of the parametric families evaluated at the matched heavy-element budget (dotted vertical line), $2.3\%$ in radius; the Vazan staircase falls within this bundle, offset from the parametric mean by $0.26\%$. \textit{(b)} The corresponding $Z(x)$ gradients, showing the range of topologies spanned by the six families.}
\label{fig:robustness_single}
\end{figure*}

\paragraph{Functional-form robustness.}
To address whether the choice of a Gaussian $Z(x)$ unduly constrains the
recovered radius, we ran the same Jupiter benchmark with six distinct
parametric families spanning the gradient topology: Gaussian, sigmoid,
exponential, bilinear, single-step, and a multi-step staircase mirroring
the layered profile of \citet{vazan_jupiters_2018}. All families share
the same envelope boundary metallicities ($Z_\mathrm{base} = 0.0816$,
$Z_\mathrm{core} = 0.240$) and the same integrated heavy-element budget,
isolating the topology of the gradient as the only varying input.
Across all six families the converged radius collapses to a mean spread of
$1.86\%$, with a maximum pairwise spread of $2.27\%$ (Fig.~\ref{fig:robustness_single}).
The digitised Vazan multi-step staircase falls within the same bundle,
offset from the parametric mean by only $0.26\%$. We conclude that, with
matched envelope boundary metallicities, the macroscopic radius is
remarkably insensitive to the exact functional topology of the gradient.
The Gaussian parameterisation adopted throughout this work is therefore a
representative, not restrictive, choice. We emphasise that this result
relies critically on the matched-boundary condition: when boundary
metallicities are allowed to drift between families, the radius spread
inflates substantially.

\subsection{Sensitivity to assumed thermal profiles}
\label{sec:thermal_robustness}

Because \texttt{fuzzycore} prescribes the temperature gradient
parametrically rather than solving energy transport self-consistently
(Sect.~\ref{sec:methods_limitations}), it is essential to demonstrate that
the converged structure is not sensitive to this choice. We tested both
the refractory core and the water mantle across the full physical range
bracketed by the isothermal, radiative-conductive, and adiabatic limits.

In the rocky core, varying the temperature gradient
$\nabla_\mathrm{core} = \mathrm{d}\ln T / \mathrm{d}\ln P$ from $0$
(isothermal) to $0.30$ (adiabatic) produces total radius variations
below $0.6\%$ across all tested configurations spanning
$0.1$--$1.0\,M_\mathrm{Jup}$ gas giants and a GJ\,1214\,b-like water
world, even as the converged central temperature varies by factors of
$1.4$--$2.7$. This reflects the pressure-dominated nature of the
iron-silicate equation of state at megabar pressures, where
$\rho(P,T) \approx \rho(P)$
\citep{seager_massradius_2007, zeng_massradius_2016}. The full sweep is
presented in Appendix~\ref{app:core_nabla}.

In the water mantle, where the equation of state retains significant
thermal expansivity through the Ice~VII / Ice~X / superionic transitions
\citep{mazevet_ab_2019}, the analogous sweep across isothermal,
polytropic, and adiabatic modes yields a more substantial response in
the mantle layer itself ($5$--$12\%$ in mantle thickness), but the total
planetary radius remains stable at the $1$--$3\%$ level, with the
largest sensitivity in water-dominated configurations. The full sweep
is presented in Appendix~\ref{app:mantle_nabla}. These results validate \texttt{fuzzycore}'s structural conclusions
against the assumed thermal profile in both refractory and volatile
layers across the physical regimes explored in this work.

\section{Applications}
\label{Sec:Application}

\subsection{Resolving the sub-Neptune degeneracy}
\label{Sec:Sub-Neptune Degeneracy}

To demonstrate the numerical stability and versatility of the framework, we applied it to two illustrative models inspired by opposite ends of the sub-Neptune spectrum: a dense, water-rich planet (a GJ\,1214\,b analogue) and a highly diffuse, gas-rich planet (a Kepler-11 e analogue). 

It must be noted that mapping an exact physical structure to observed mass and radius is a highly degenerate inverse problem requiring rigorous parameter space exploration. Therefore, these models are presented as structural analogues. This is better explored in Sect.~\ref{sec:atlas}.

\subsubsection{The volatile-rich analogue: GJ\,1214\,b-like}

GJ\,1214\,b is the archetypal volatile-rich sub-Neptune. It was discovered in transit by the
MEarth survey \citep{charbonneau_super-earth_2009}, and its mass has since been pinned down by
radial velocity: a decade-long HARPS campaign of 165 velocities, analysed jointly with archival
transits, gives $M_\mathrm{p} = 8.17 \pm 0.43\,M_\oplus$ and
$R_\mathrm{p} = 2.742^{+0.050}_{-0.053}\,R_\oplus$ \citep{cloutier_more_2021}. The corresponding
bulk density, $\rho \approx 2.2$\,g\,cm$^{-3}$, lies far below that of any plausible rock-iron
body of this mass and therefore demands a substantial volatile component, whether as a deep
water layer, an extended H/He envelope, or some combination of the two. That ambiguity is
precisely the degeneracy our framework is built to map, and it is why we adopted GJ\,1214\,b as our
volatile-rich test case.

The host is a nearby mid-M dwarf ($M_\star = 0.178 \pm 0.010\,M_\odot$,
$R_\star = 0.215 \pm 0.008\,R_\odot$, $d = 14.63$\,pc; \citealt{cloutier_more_2021}). Its small
radius yields a large transit depth and makes the planet the most favourable sub-Neptune target
for transmission spectroscopy; panchromatic JWST observations accordingly indicate a highly
metal-enriched atmosphere or a thick aerosol layer \citep{kempton_reflective_2023}. For our purposes the stellar type enters only through the atmospheric boundary condition: the temperature at the top of the integration domain, which we impose at $1$\,bar and prescribe rather than derive self-consistently. It does not otherwise alter the interior equations we solve.

While traditional integrators struggle with boundary matching in such high-metallicity
environments, our framework explicitly handles a high-$Z$ boundary condition, allowing a
seamless transition between a dense envelope and a supercritical mantle. The planet is modelled
at the measured mass with a $6.55 \,M_{\oplus}$ mixed core ($33\%$ iron, $67\%$ rock), a
condensed $1.57 \,M_{\oplus}$ water mantle, and a thin, highly enriched $0.05 \,M_{\oplus}$
envelope. Because the vast majority of the mass resides in dense, highly incompressible phases,
the physical radius remains compact: the solver drives the central pressure to $6.51$\,Mbar, a
realistic value for an $8 \,M_{\oplus}$ rock-water body, and converges on a total radius of
$2.72 \,R_{\oplus}$ (see Appendix~\ref{app:profiles} for the full internal profiles). This sits
within $1\%$ of the measured radius, demonstrating that a water-rock interior with a
compositional gradient can reproduce the observation. The structural parameters are summarised
in Table~\ref{table:subneptunes}. We note, however, that an adiabatic gradient would not be
expected all the way up to $1$\,bar; this is examined in Sect.~\ref{sec:HADES} and motivates the
use of coupled atmosphere--interior models.

\subsubsection{The dilute super-puff analogue: Kepler-11 e-like}

Kepler-11 e sits at the opposite extreme. It is one of six transiting planets in a tightly
packed system discovered by \textit{Kepler} \citep{lissauer_closely_2011}. Unlike GJ 1214, the host is too faint ($V = 14.2$\,mag) for a precision radial-velocity mass; the planetary masses are instead
dynamical, derived from the transit-timing variations induced by mutual perturbations within
the compact system. An analysis of 40 months of \textit{Kepler} photometry gives $M_\mathrm{p} = 8.0^{+1.5}_{-2.1}\,M_\oplus$ and $R_\mathrm{p} = 4.19^{+0.07}_{-0.09}\,R_\oplus$ \citep{lissauer_all_2013}, and the host has since been reclassified as a young solar twin ($M_\star \approx 1.04\,M_\odot$),  a revision that raised the inferred planetary densities by $20$--$95\%$ \citep{bedell_kepler-11_2017}. The revised stellar density is, however, in $\sim 2\sigma$ tension with the value obtained from the transit and timing data, including in \citeauthor{bedell_kepler-11_2017}'s own photodynamical fit, which favours a stellar density closer to that of \citet{lissauer_all_2013}. We therefore adopted the \citet{lissauer_all_2013} planetary parameters throughout.

Even so, the resulting bulk density, $\rho = 0.58^{+0.11}_{-0.16}$\,g\,cm$^{-3}$, is roughly a quarter of
GJ\,1214\,b's at almost the same planetary mass. The two objects therefore bracket the
sub-Neptune population at fixed $M_\mathrm{p} \approx 8\,M_\oplus$: one compact and
volatile-dominated, one extended and gas-dominated. That contrast is what motivates their
selection here. As in the GJ\,1214\,b case, the host star enters our calculation only through the temperature
imposed at the $1$\,bar boundary, which we prescribe rather than derive self-consistently (Table~\ref{table:subneptunes}). The coupled atmosphere--interior calculation of Sect.~\ref{sec:HADES} illustrates how this boundary condition is obtained in a fully self-consistent
treatment.

To model this object we allocated a $0.33 \,M_{\oplus}$ gaseous envelope over a
$7.67\,M_{\oplus}$ core anchor, omitting a distinct condensed water layer. If that envelope
were composed of pure, fully convective hydrogen and helium resting on a sharp core boundary,
the resulting planet would be far larger than observed. To fit such a gas fraction into
Kepler-11 e's compact volume, the envelope must be heavily polluted.

By applying our adaptive atmospheric modelling and solving for a wide compositional gradient
within the envelope ($\sigma = 0.20$), the framework reconciles this discrepancy. The dilute
heavy elements lofted into the envelope drastically increase the mean molecular weight ($\mu$)
of the deep gas. The resulting rise in density, and the corresponding fall in atmospheric scale
height, physically compacts the bloated atmosphere. This `$\mu$-compaction' collapses the
envelope down to a total radius of $4.17 \,R_{\oplus}$ (Fig.~\ref{fig_kepler11e_app}), within
$0.4\%$ of the measured value.

It is important to note a critical caveat regarding super-puffs: while our model demonstrates
that a deeply suspended heavy-element gradient can ballast a massive envelope down to the
required radius, apparent extreme low densities may also be driven by high intrinsic
temperatures associated with youth, escaping hydrodynamic winds, or optically thick
photochemical hazes \citep{rappaport_possible_2012, gaidos_zodiacal_2017}. These non-hydrostatic
scenarios offer compelling competing or complementary explanations for the observed volume.

\begin{table}[t]
\caption{Contrasting interior models for sub-Neptune analogues.}
\centering
\resizebox{\columnwidth}{!}{
\begin{tabular}{l c c c}
\hline\hline
Parameter & GJ\,1214\,b-like & Kepler-11 e-like & Unit \\
\hline
Top temperature (in)      & 500.0 & 900.0 & K \\
Top pressure (in)         & 1.0   & 1.0   & bar \\
Top Z fraction (in)       & 0.05  & 0.02  & $-$ \\
\hline
Core mass ($33\%$ Fe) (in) & 6.55 & 7.67    & $M_{\oplus}$ \\
Water mantle mass (in)     & 1.57 & \ldots  & $M_{\oplus}$ \\
Envelope mass (in)         & 0.05 & 0.33    & $M_{\oplus}$ \\
Staircase $\sigma$ (in)    & 0.10 & 0.20    & $-$ \\
Total mass (converged)     & 8.12 & 8.00    & $M_{\oplus}$ \\
\hline
$P_\mathrm{c}$             & 6.51 & 6.54    & Mbar \\
$R_\mathrm{rock}$          & 1.92 & 2.04    & $R_{\oplus}$ \\
$R_\mathrm{water}$         & 2.39 & \ldots  & $R_{\oplus}$ \\
$R_\mathrm{total}$         & 2.72 & 4.17    & $R_{\oplus}$ \\
\hline
Observed $R_\mathrm{p}$    & 2.74 & 4.19    & $R_{\oplus}$ \\
\hline
\end{tabular}
}
\tablefoot{Observed radii from \citet{cloutier_more_2021} and \citet{lissauer_all_2013}.}
\label{table:subneptunes}
\end{table}

The successful convergence on both targets highlights the robustness of the nested shooting method. The framework smoothly manages the transition from the nearly negligible atmospheric interface of the GJ\,1214\,b analogue to the deep, high-pressure boundary beneath the massive, metal-polluted envelope of the Kepler-11 e analogue, validating its unified approach to sub-Neptune degeneracy.

\subsubsection{Mean molecular weight compaction and thermal blanketing}
The profound structural impact of redistributing heavy elements is best visualised by tracking the competing physical forces within a planet as its core becomes increasingly fuzzy. To evaluate this we propose two studies. On one hand, a direct mass transfer of core to envelope at a fixed dilution width ($\sigma$) named `bulk envelope transfer', and on the other, a direct metal transfer, `direct metal', where $\sigma$ is adjusted to strictly transfer core mass into dilute heavy element (metal) mass without altering the total planetary mass or bulk metallicity. The latter conserves the total heavy-element budget ($M_Z$) to better than $0.02\,M_\oplus$ in all reported cases.

To rigorously converge on these required planetary structures, the framework employs a hybrid numerical root-finding scheme. Initially, a coarse bisection search is used to safely map the parameter space and secure a valid physical bracket. Once a safe boundary is established, the algorithm switches to Brent's method. This transition leverages the local slope of the error function to achieve rapid, high-precision convergence, combining absolute physical safety with computational speed.

\begin{figure}[t!]
\centering
\includegraphics[width=\columnwidth]{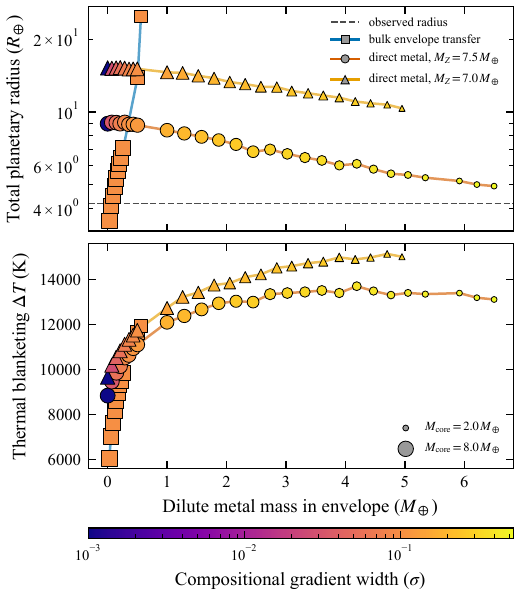}
\caption{1D parameter sweep for a Kepler-11 e-like planet at a fixed total mass ($M_\mathrm{tot} = 7.95\,M_\oplus$). \textit{Top panel:} competing structural effects of envelope enrichment. The bulk envelope transfer track inflates the planet by replacing core mass with H/He and metals, while the direct metal tracks demonstrate the transition from atmospheric expansion to $\mu$-compaction. \textit{Bottom panel:} thermodynamic consequence of dilute cores. Regardless of the geometric response, metal-rich envelopes act as a thermal blanket, increasing the temperature difference between surface and core. Marker colour encodes the dilution width $\sigma$, marker size the solid core mass; the dashed horizontal line marks the observed radius of Kepler-11 e \citep{lissauer_all_2013}.}
\label{fig_kepler11e_sweep}
\end{figure}

As shown in the bulk envelope transfer track of Fig.~\ref{fig_kepler11e_sweep}, there is a monotonic increase in planetary radius as solid core mass is reduced. In this mode, the dense core is replaced by a mixture of H/He gas and heavy elements within the envelope. The addition of highly compressible hydrogen and helium, combined with the volumetric expansion of the atmosphere, leads to a runaway inflation that far exceeds the radius of a standard core-envelope model.

Conversely, the direct metal tracks, which strictly conserve the total heavy element budget ($M_Z$), reveal that $\mu$-compaction dominates the radius response across nearly the entire physically relevant range. A very narrow initial geometric-expansion phase is present at $M_\mathrm{dilute} \lesssim 0.3\,M_\oplus$, where the envelope swells slightly under the added atmospheric mass, but this rise contributes less than $0.15\,R_\oplus$ in absolute terms before the structural response inverts. Beyond this threshold, the rising mean molecular weight ($\mu$) of the metal-enriched envelope dominates: the local density and gravitational binding of the gas increase, and the envelope is physically crushed inwards. The direct metal tracks therefore exhibit a near-monotonic radius compaction over their dilute-mass range, with the displayed $M_Z = 7.0$ and $M_Z = 7.5\,M_\oplus$ tracks contracting by $32\%$ and $45\%$ respectively as the envelope metal content rises from zero to its maximum.

Beyond these geometric changes, the redistribution of heavy elements profoundly alters the planet's thermal profile (Fig.~\ref{fig_kepler11e_sweep}, bottom panel). By isolating the difference in temperature between top and bottom of the envelope, we observe a severe, near-monotonic rise in the temperature at the core-envelope interface as the envelope becomes enriched: $\Delta T$ increases by approximately $5000\,\mathrm{K}$ (a factor of $\sim 1.5$) across the explored range. The bulk envelope transfer track is strictly monotonic, while the direct metal tracks rise monotonically through the bulk of the parameter space before plateauing at the highest dilute-mass fractions. In the bulk envelope transfer track, this increase is driven by the massive addition of H/He gas, which vastly increases the envelope's total heat capacity. Crucially, the direct metal tracks demonstrate a pure thermal blanketing effect: as heavy elements are vaporised into the envelope, the resulting increase in atmospheric opacity and the introduction of double-diffusive entropy jumps lead to a redistribution of heat closer to the core. Regardless of whether the planet geometrically expands marginally or undergoes substantial $\mu$-compaction, the metal-rich envelope acts as a strict thermal throttle.

This reveals a fundamental physical constraint: when bulk metallicity is strictly conserved, redistributing metals into the envelope does not structurally inflate the planet. Rather, $\mu$-compaction actively suppresses thermal expansion. This is a purely static structural finding: at fixed total heavy-element
budget, redistributing metals from a compact core into a dilute envelope
does not inflate the planet's hydrostatic radius. The temperature contrast
between surface and core that accompanies this redistribution is a
diagnostic of the converged static profile, not a measurement of cooling
behaviour; whether such a planet would in practice retain or shed
primordial heat over geological time is a question for evolutionary
modelling (Sect.~\ref{sec:landscape}, Sect.~\ref{sec:radiative_layers}) and lies
outside the scope of the static framework.

The same structural reasoning constrains the family of static interiors
consistent with extreme low-density objects (e.g.\ super-puffs): an
interior architecture in which heavy elements are lofted throughout an
extended envelope incurs a $\mu$-compaction penalty that is mechanically
incompatible with the very low bulk densities observed. Pristine,
low-metallicity envelopes are therefore structurally favoured for these
objects, with alternative inflation pathways, intense tidal heating
\citep{millholland_tidally_2019}, ongoing hydrodynamic escape
\citep{rappaport_possible_2012}, or optically thick high-altitude hazes
\citep{gao_aerosol_2020}, providing the additional non-hydrostatic
mechanisms required to bridge the remaining radius gap.

\subsection{Internal structure and compositional trajectories}
\label{sec:trajectories}

\begin{figure*}[ht!]
\centering
\includegraphics[width=0.95\textwidth]{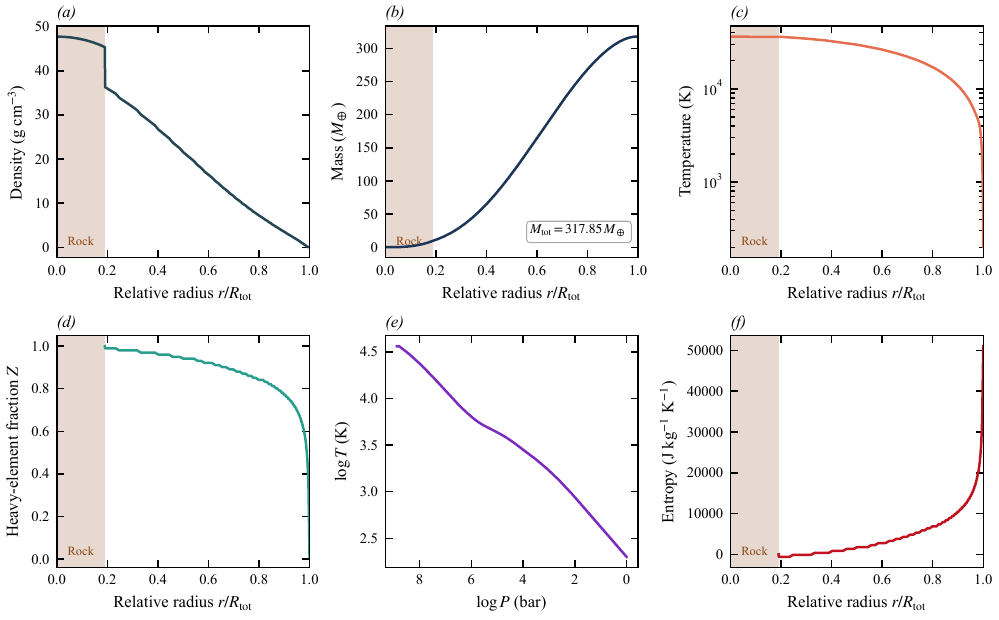}
\caption{Internal structure of a $1.0\,M_\mathrm{Jup}$ model with a dilute compositional gradient ($\sigma = 0.05$). \textit{(a)} density,
\textit{(b)} enclosed mass, \textit{(c)} temperature, \textit{(d)} the
staircase heavy-element (water) fraction $Z$, \textit{(e)} the resulting $P$--$T$ adiabat, and \textit{(f)} specific entropy. All panels except (e) are plotted against relative radius; the brown shaded region marks the solid core.}
\label{fig_diagnostics}
\end{figure*}

The fundamental design of our model is best illustrated by examining the internal profiles of a representative giant planet ($1.0\,M_\mathrm{Jup}$, $T_{\text{P=1bar}} = 200$\,K). In Fig.~\ref{fig_diagnostics}, we present this model featuring a dilute compositional gradient ($\sigma = 0.05$), which yields a converged radius of $0.87\,R_\mathrm{Jup}$. To isolate the structural impact of this heavy-element distribution, we provide a fully convective, gradient-free baseline model in Appendix~\ref{app:jupiter_profile}. Contrasting the gradient-bearing model with the uniform baseline (which yields a radius of $0.97\,R_\mathrm{Jup}$) physically demonstrates how the presence of the dilute region and its associated entropy jumps alter the deep interior profile and the resulting global radius. The detailed thermodynamic trajectory of this dilute-core model across the 3D mixed EOS phase space is provided in Appendix~\ref{app:phase_space}.

\subsection{The mass-temperature equivalence grid}
\label{sec:mass_temp_grid}

To characterise the degeneracies of the fuzzy core model, we produced a $3 \times 3$ grid of structural equivalence maps. This grid explores three representative planetary masses ($0.3, 1.0,$ and $5.0\,M_\mathrm{Jup}$) across three boundary temperatures at $1$\,bar ($200$, $500$, and $1000$\,K).

\begin{figure*}[ht!]
\centering
\includegraphics[width=0.95\textwidth]{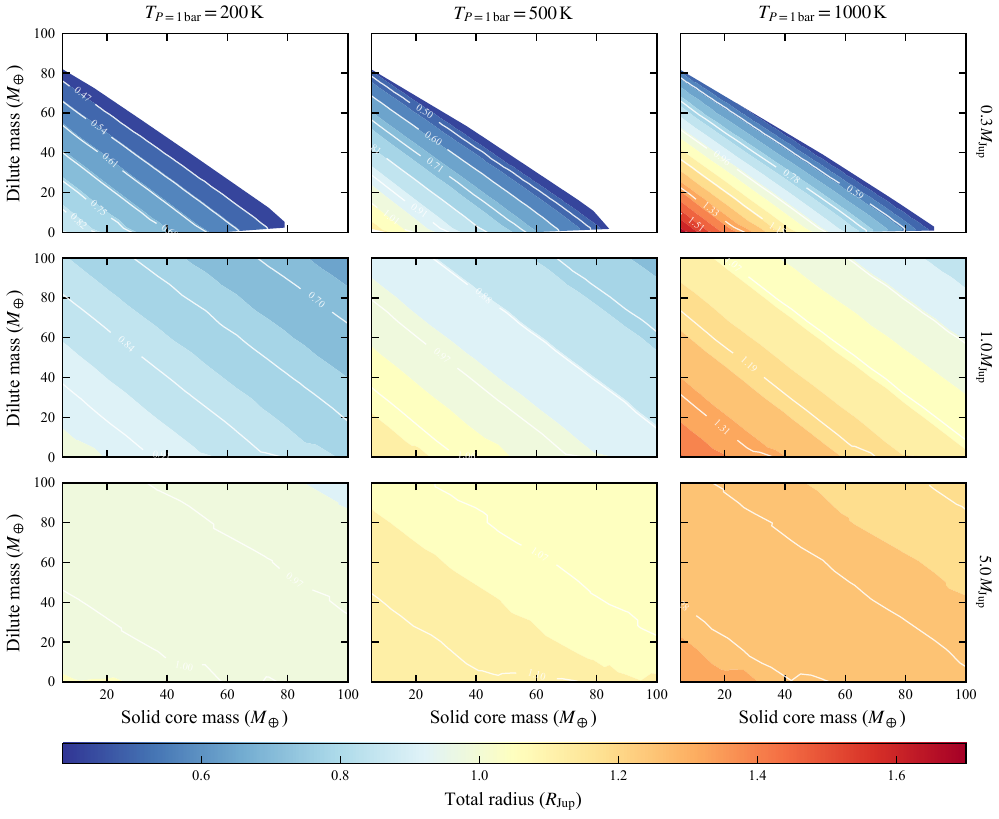}
\caption{Structural equivalence grid for giant planets. Rows represent total planetary mass ($0.3$, $1.0$, $5.0\,M_\mathrm{Jup}$), and columns the temperature at $1$\,bar ($200$, $500$, $1000$\,K). The white contours define iso-radius paths in the $M_\mathrm{core}$--$M_\mathrm{dilute}$ plane, demonstrating the structural trade-off between concentrated and distributed heavy elements. The iso-radius contours are linearly spaced within the radius range converged in each panel, chosen to expose the curvature of the equivalence surface rather than to match any specific observed object.}
\label{fig_grid}
\end{figure*}

As illustrated in Fig.~\ref{fig_grid}, the role of internal heavy element distribution varies significantly with total planetary mass. In the low-mass regime ($0.3\,M_\mathrm{Jup}$), the core is highly degenerate. There exists a clear equivalence between placing metals in a solid core versus a dilute gradient. The iso-radius curves show a pronounced negative slope, indicating that to maintain a constant planetary radius, an increase in dilute mass must be directly offset by a decrease in solid core mass. However, the solid core plays a powerful structural role; concentrating metals deep in the gravity well shrinks the total radius much more efficiently than lofting them into the envelope.

The structural trade-offs are strongly mediated by the planet's thermal state. In warmer planets ($1000$\,K), the solid core plays a progressively greater role in determining the final radius. This occurs because the solid materials (iron and rock) expand significantly less under thermal forcing than the gaseous H/He envelope. Therefore, sequestering heavy elements into a dense, thermally resistant core is much more effective at keeping the planet compact than mixing them into a hot, highly expansive dilute envelope. This is reflected in the steepening of the iso-radius contours as temperature increases.

As total planetary mass increases to $5.0\,M_\mathrm{Jup}$, the structural dynamics shift primarily in their magnitude of response. When zooming into physically realistic heavy element budgets (e.g. up to $100\,M_{\oplus}$ of heavy elements), we observe that the iso-radius lines remain clearly diagonal. This indicates that the fundamental equivalence, the structural trade-off between compact solid core mass and distributed dilute mass, persists even in extreme gravity environments. However, the total variation in planetary radius across this parameter space is drastically reduced. At these extreme internal pressures, the massive $5.0\,M_\mathrm{Jup}$ H/He envelope becomes partially degenerate and deeply self-compressed. Because $100\,M_{\oplus}$ of metals represents a very small fraction of the total planetary mass, its addition acts only as a minor perturbation to the bulk density. Even with $100\,M_\oplus$ of metals distributed across a wide
compositional staircase, the integrated radius change in these high-mass
configurations remains below the per-cent level. Whether such a staircase
would survive against turbulent entrainment over geological time, and
the corresponding consequence for the cooling history of the planet,
falls to evolutionary modelling
\citep{vazan_jupiters_2018, sur_apple_2024, helled_giant_2025}.

\subsection{Water-world degeneracy atlas}
\label{sec:atlas}

The local degeneracies illustrated by the GJ\,1214\,b and Kepler-11 e
analogues (Sect.~\ref{Sec:Sub-Neptune Degeneracy})
are pedagogical: each model occupies a single point in a much larger
continuous family. To map this family at the resolution evolutionary codes
cannot practically reach, we deployed \texttt{fuzzycore} across a
$15 \times 15 \times 12$ grid in the water-mass, envelope-mass, and
dilution-width plane ($M_\mathrm{water} \in [0, 4]\,M_\oplus$,
$M_\mathrm{env} \in [0.05, 3]\,M_\oplus$, $\sigma \in [10^{-3}, 0.6]$
logarithmically spaced) at fixed total mass $M_\mathrm{p} = 8\,M_\oplus$. Of the
$2700$ grid points, $2155$ converged ($80\%$ success rate). 

As detailed in Appendix~\ref{app:failure_topology}, the non-converging models are not randomly distributed but exhibit a distinct failure topology. We observe a strong concentration of solver failures in regimes characterised by low dilution widths ($\sigma \lesssim 0.01$) combined with high water mantle masses ($M_\mathrm{water} \to 4.0\,M_\oplus$). In these extreme architectures, the gaseous envelope becomes barometrically ultra-thin; attempting to resolve a highly compressed, sharp compositional gradient within this shallow domain pushes the adaptive stepper to its limits, occasionally triggering numerical instabilities in the multi-phase EOS interpolator. Consequently, this localised breakdown translates into a moderate baseline failure rate for the isolated high-water and low-$\sigma$ subsets. 

We note that the grid in $(M_{\mathrm{water}}, M_{\mathrm{env}}, \sigma)$ is bounded only by the requirement of hydrostatic convergence; consistency of the resulting $(P_{\mathrm{int}}, T_{\mathrm{int}}, Z_{\mathrm{core}})$ with the water--H/He demixing critical curve is not enforced (Sect.~\ref{sec:methods_limitations}). The atlas should therefore be read as the set of mechanically permitted static configurations, a superset of the thermodynamically self-consistent ones, and the relevant prior for any subsequent retrieval or evolutionary follow-up.

\begin{figure*}[ht!]
\centering
\includegraphics[width=0.95\textwidth]{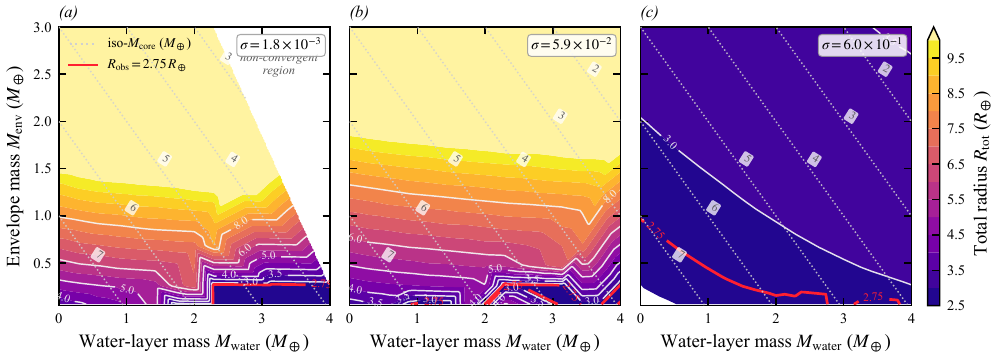}
\caption{Radius landscape of static sub-Neptune interiors at fixed
$M_\mathrm{p} = 8\,M_\oplus$, computed across a $15 \times 15 \times 12$ grid in $(M_\mathrm{water}, M_\mathrm{env}, \sigma)$ with $2155$ converged models. Panels show the converged $R_\mathrm{tot}$ over $(M_\mathrm{water}, M_\mathrm{env})$ at three representative dilution widths: \textit{(a)} sharp, \textit{(b)} moderate, and \textit{(c)} diffuse. The value of $\sigma$ is given in each panel. Dotted diagonals trace constant rocky-core mass ($M_\mathrm{core} = M_\mathrm{p} - M_\mathrm{water} - M_\mathrm{env}$), labelled in $M_\oplus$. The red contour marks $R_\mathrm{obs} = 2.75\,R_\oplus$ in all three panels, demonstrating that the iso-radius locus drifts substantially with $\sigma$: the dilution width is an independent and observationally degenerate dimension at fixed
$(M_\mathrm{p}, R_\mathrm{p})$.}
\label{fig:atlas}
\end{figure*}

For the successfully converged models, the radii span $2.4\,R_\oplus$ (compact, sharp-boundary low-envelope cases) to $19.7\,R_\oplus$ (puffed-up sharp-boundary high-envelope cases). Fig.~\ref{fig:atlas} presents three representative $\sigma$ slices through this volume. At sharp dilution ($\sigma = 0.002$), the radius response is steep in $M_\mathrm{env}$: a single Earth mass of H/He on modest core+water inflates the planet beyond $10\,R_\oplus$. At moderate dilution ($\sigma = 0.059$), $\mu$-compaction begins to suppress this inflation across the upper part of the plane. At diffuse dilution ($\sigma = 0.60$), $\mu$-compaction dominates entirely: the entire $(M_\mathrm{water}, M_\mathrm{env})$ plane is compressed below $4\,R_\oplus$. The red contour marks the locus of $R_\mathrm{obs} = 2.75\,R_\oplus$ (a GJ\,1214\,b-like radius) across all three panels, drifting from a tight band at low $M_\mathrm{env}$ at sharp $\sigma$ to a broad sweep spanning the full $M_\mathrm{water}$ axis at diffuse $\sigma$.

The structural consequence is striking: at fixed observed $(M_\mathrm{p}, R_\mathrm{p})$,
the dilution width $\sigma$ is essentially unconstrained. Of the $88$
interior solutions consistent with $R_\mathrm{obs} = 2.75 \pm 0.10\,R_\oplus$,
all $12$ sampled values of $\sigma$ are represented, as are $14$ of the
$15$ sampled values of $M_\mathrm{water}$. Atmospheric metallicity priors
from JWST or \textit{Ariel} can constrain $Z_\mathrm{base}$ and therefore restrict
parts of this family; the dilution width itself, however, leaves no
direct atmospheric signature and remains observationally degenerate
without additional gravity-field information or independent core-mass
priors. The atlas thus quantifies the irreducible ambiguity of static
sub-Neptune characterisation from $(M_\mathrm{p}, R_\mathrm{p})$ alone.

\section{Discussion}
\label{sec:discussion}

\subsection{\texttt{fuzzycore} in the broader modelling landscape}
\label{sec:landscape}
As demonstrated in the preceding sections, the specific spatial distribution of heavy elements within a planetary interior introduces profound mass-radius degeneracies. Addressing these degeneracies has driven the adaptation of stellar evolutionary codes (Henyey solvers) to planetary physics. Frameworks such as \texttt{MESA} \citep{helled_giant_2025}, \texttt{CEPAM} \citep{guillot_giant_2015}, and \texttt{APPLE} \citep{sur_apple_2024} track the self-consistent evolution of planetary interiors over gigayears. These state-of-the-art codes dynamically couple the equations of structure, radiative-convective energy transport, and compositional mixing, allowing researchers to study emergent, time-dependent phenomena such as the erosion of primordial gradients \citep{vazan_jupiters_2018}, the formation of semi-convective layers, and helium phase separation \citep{vazan_evolution_2016}.

The contribution of \texttt{fuzzycore} is not to supersede the microphysical rigour or time-dependent capabilities of these evolutionary frameworks. Rather, it is designed to be highly complementary. Because fully implicit Henyey solvers must invert complex Jacobian matrices and strictly conserve energy over billions of years, they are computationally expensive. This expense often prohibits the dense, large-scale forward modelling required to map exact iso-radius contours across thousands of possible atmospheric metallicities, core mass fractions, and gradient topologies. 

By decoupling static structural profiling from temporal evolution and dynamic mixing, \texttt{fuzzycore} acts as a rapid parameter-space explorer. Its niche is to quickly identify the specific families of static architectures that are mechanically permitted by a given $(M_\mathrm{p}, R_\mathrm{p})$ observation at a fixed snapshot in time. Once the degenerate volume is rapidly mapped by a static solver, targeted evolutionary codes can then be deployed to ascertain whether those specific permitted architectures could realistically form and survive over the age of the system.

\subsection{Radiative layers and fuzzy-core survival}
\label{sec:radiative_layers}
A critical limitation of the static snapshot approach is its inability to self-consistently determine the boundaries of radiative and convective zones based on local opacities. In our framework, the temperature gradient ($\nabla$) is prescribed parametrically to evaluate the structural sensitivity of the planet to different thermal profiles (as demonstrated in Appendix~\ref{app:core_nabla}).

However, the nature of the temperature gradient is intrinsically linked to the physical survival of the compositional gradient itself. Recent evolutionary models demonstrate that heavy-element gradients often suppress convection entirely, leading to the formation of deep radiative layers \citep{knierim_convective_2024, arevalo_jupiter_2024}. In these radiative zones, mixing is restricted to microscopic diffusion, meaning the fuzzy core can survive effectively un-eroded for billions of years \citep{sur_next-generation_2026, tejada_arevalo_sub-neptune_2026}. Conversely, if the deep interior remains convective (e.g. due to a `hot start' formation or high internal heat flux), turbulent entrainment can mix the heavy elements, eroding the fuzzy core and homogenising the envelope \citep{vazan_jupiters_2018}. 

Because \texttt{fuzzycore} does not evaluate energy transport or mixing timescales, it cannot predict whether a given Gaussian or staircase profile is hydrodynamically stable. The structural atlases generated by this code must therefore be interpreted as maps of mechanically stable states, while the ultimate viability of those states relies on the long-term radiative-convective history modelled by time-dependent codes.

\subsection{Implications for super-puffs and dilute envelopes}
The structural competition between thermal expansion and mean molecular weight ($\mu$) compaction highlighted in this framework provides a useful mechanical lens for interpreting super-puffs, planets with anomalously low bulk densities \citep[e.g. Kepler-51;][]{masuda_very_2014}. When heavy elements are concentrated in a compact central core, the overlying H/He envelope possesses a minimum mean molecular weight, maximising its compressibility and scale height. While massive energy dissipation mechanisms (such as intense tidal heating or retained primordial heat) are frequently invoked to inflate these planets \citep{millholland_tidally_2019}, our framework demonstrates that the spatial distribution of the metals is equally critical. If the heavy elements are lofted into the envelope to form a diffuse fuzzy core, the resulting $\mu$-compaction physically suppresses the thermal expansion. Therefore, if a super-puff is observed to have a highly enriched atmosphere via transmission spectroscopy, its extreme radius becomes exceedingly difficult to reproduce statically, as the structural compaction of the metals fights the thermal inflation required by the low density.

\subsection{Atmosphere-interior coupling and spectral forward modelling}
\label{sec:HADES}
The structural equivalence grids generated in our water-world atlas
highlight that planetary mass and total radius alone are profoundly
insufficient to constrain interior architectures. Breaking this
degeneracy requires directly linking the deep interior to observable
atmospheric spectra to provide boundary priors on the envelope
metallicity.

To demonstrate this capability, \texttt{fuzzycore} is natively
structured to couple with external atmospheric radiative transfer
modules, such as the Heat Atmosphere Density Evolution Solver
\citep[\texttt{HADES};][]{wilkinson_breaking_2024}. In a coupled forward-modelling
architecture, the atmospheric module calculates the line-by-line
radiative-convective equilibrium of the optically thin upper
atmosphere, providing the boundary conditions ($P_{\text{surf}},
T_{\text{surf}}$) to the deep interior. Because \texttt{fuzzycore} can
dynamically solve for a target surface gravity ($g$) via its nested
shooting method, it returns a self-consistent deep interior
architecture that matches the gravitational constraints of the
overlying atmosphere.

\begin{figure*}[ht!]
\centering

\begin{minipage}[c]{0.46\textwidth}
  \centering
  \includegraphics[width=\linewidth]{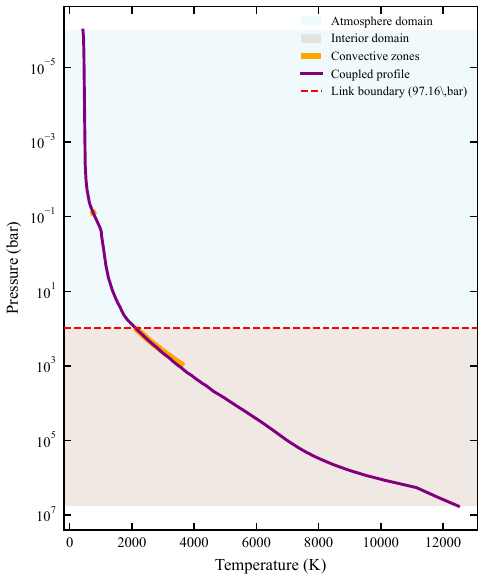}
\end{minipage}
\begin{minipage}[c]{0.46\textwidth}
  \centering
  \includegraphics[width=\linewidth]{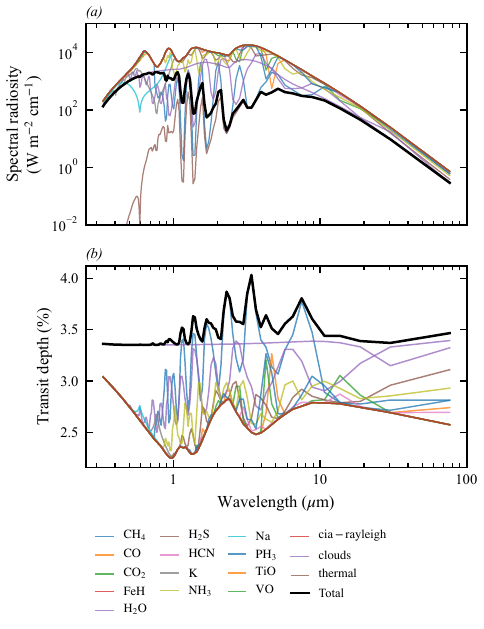}\\
\end{minipage}

\caption{Coupled atmosphere--interior model for a GJ\,1214\,b analogue computed by \texttt{fuzzycore} within the \texttt{HADES} radiative-convective framework, with $\sigma = 0.34$, $M_\mathrm{core} = 3.7\,M_\oplus$, $M_\mathrm{water} = 4\,M_\oplus$, and $T_\mathrm{int} = 150$\,K. \textit{Left:} continuous $P$--$T$ profile spanning the optically thin upper atmosphere through the deep envelope and condensed water mantle; the dashed red line marks the pressure at which the atmospheric and interior modules are linked, and the shaded bands indicate the two domains. \textit{Right:} the resulting synthetic spectra, \textit{(a)} thermal emission and \textit{(b)} transmission, each decomposed into the contributions of the individual molecular and aerosol species.}
\label{fig:hades_combined}
\end{figure*}

Fig.~\ref{fig:hades_combined} illustrates this coupling for a single
GJ\,1214\,b analogue drawn from the water-world atlas
(Sect.~\ref{sec:atlas}). The left panel shows the continuous $P$--$T$
profile spanning the optically thin upper atmosphere through the
deep envelope and condensed water mantle, computed self-consistently
across both modules. The right panels present the resulting synthetic
emission (top) and transmission (bottom) spectra, with the
contributions of each molecular and aerosol species shown
explicitly. The decomposition makes the structural-to-spectral mapping
diagnostic: each chemical and cloud contributor leaves a distinct
signature whose amplitude depends on the converged $(P, T, Z)$ profile
returned by \texttt{fuzzycore}. We present this single configuration
as a proof of concept rather than a converged retrieval; a full
coupled sweep across the atlas of Sect.~\ref{sec:atlas} is the natural
next step and is deferred to future work.

This demonstration confirms that \texttt{fuzzycore} can serve as a
static backend for atmospheric retrievals, allowing the high-accuracy
spectroscopic constraints anticipated from JWST and the upcoming ESA
\textit{Ariel} mission \citep{tinetti_atmospheric_2020} to be directly mapped
onto deep interior architectures.

\section{Conclusions}
\label{sec:conclusions}

In this work, we introduced \texttt{fuzzycore}, a rapid, multi-zone hydrostatic solver designed to map the profound structural degeneracies introduced by compositionally graded planetary interiors. Our primary findings and the capabilities of the framework are summarised as follows:

\begin{enumerate}
    \item A unified static solver: \texttt{fuzzycore} successfully links state-of-the-art ab initio EOS tables for iron-rock cores, condensed volatile mantles, and compositionally graded envelopes within a single, open-source numerical framework. 
    \item Static evolutionary benchmarking: When initialised with the exact mass coordinates and heavy-element distribution of an evolution-derived Jupiter model \citep{vazan_jupiters_2018}, the static solver converges to a macroscopic radius within $\sim 10\%$ of the true planet, with the residual gap primarily attributable to documented differences in the underlying equations of state.
    \item Functional-form robustness: We empirically demonstrate that when the integrated heavy-element budget and the envelope boundary metallicities are strictly matched, the converged macroscopic radius of a giant planet is remarkably insensitive to the exact functional topology of the compositional gradient. Across six distinct parametric families (including a complex hydrodynamic staircase), the resulting radii collapse to within a $2.3\%$ spread. This demonstrates that smooth analytical parameterisations, such as a Gaussian decay, are representative, non-restrictive choices for mapping macroscopic degeneracies.
    \item Mapping the water-world degeneracy: By deploying the solver across the sub-Neptune regime, we generated a dense structural atlas for a GJ\,1214\,b analogue. The atlas reveals a continuous family of static architectures, trading condensed water mass against envelope metallicity and gradient width ($\sigma$), that all reproduce the observed $(M_\mathrm{p}, R_\mathrm{p})$. While atmospheric metallicity priors from JWST can constrain this space, a significant degenerate volume remains inherently unresolvable without additional physical priors.
    \item Complementary capability: By stripping away the computational overhead of time-dependent energy transport and convective mixing, \texttt{fuzzycore} complements fully implicit Henyey evolutionary codes by enabling the dense, high-resolution static mapping of parameter spaces that are otherwise too computationally expensive to comprehensively sweep.
\end{enumerate}

\begin{acknowledgements}
We thank the anonymous reviewers for their constructive comments, which helped improve the quality of this manuscript. This project has received funding from the European Research Council (ERC) under the European Union’s Horizon 2020 research and innovation programme (COBREX; grant agreement n° 885593). This work was granted access to the HPC resources of MesoPSL financed by the Region Ile de France and the project Equip@Meso (reference ANR-10-EQPX-29-01) of the programme Investissements d’Avenir supervised by the Agence Nationale pour la Recherche. This work was supported by CNES, focused on AIRS on \textit{Ariel}, and by the Programme National de Planétologie (PNP) of CNRS/INSU, cofunded by CNES. The authors acknowledge the use of Google's Gemini (version 3.1 Pro, \url{https://gemini.google.com}) for assistance with code development, brainstorming, language editing, and grammatical polishing. Special thanks to the open-source community for the SciPy \citep{virtanen_scipy_2020} and NumPy libraries. The Python frameworks developed and utilised in this study, \texttt{fuzzycore} and \texttt{HADES}, are fully open-source and available on GitHub at \url{https://github.com/ChristianSWilkinson/fuzzycore} and \url{https://github.com/ChristianSWilkinson/exoweave}, respectively. The repositories include the nested boundary-shooting solver, the atmospheric radiative transfer modules, and the execution scripts required to reproduce the parameter sweeps and coupled structural models presented in this work.
\end{acknowledgements}

\bibliographystyle{aa} 
\bibliography{library} 

\begin{appendix}
\renewcommand{\topfraction}{0.95}
\renewcommand{\bottomfraction}{0.95}
\renewcommand{\textfraction}{0.02}
\renewcommand{\floatpagefraction}{0.6}
\setcounter{topnumber}{4}
\setcounter{totalnumber}{6}
\setcounter{dbltopnumber}{3}
\renewcommand{\dbltopfraction}{0.95}
\renewcommand{\dblfloatpagefraction}{0.6}
\section{Supplementary figures and sensitivity tests}
\label{app:supplementary}

\subsection{Internal structure profiles of sub-Neptune analogues}
\label{app:profiles}

\begin{figure*}[ht!]
\centering
\includegraphics[width=0.7\textwidth]{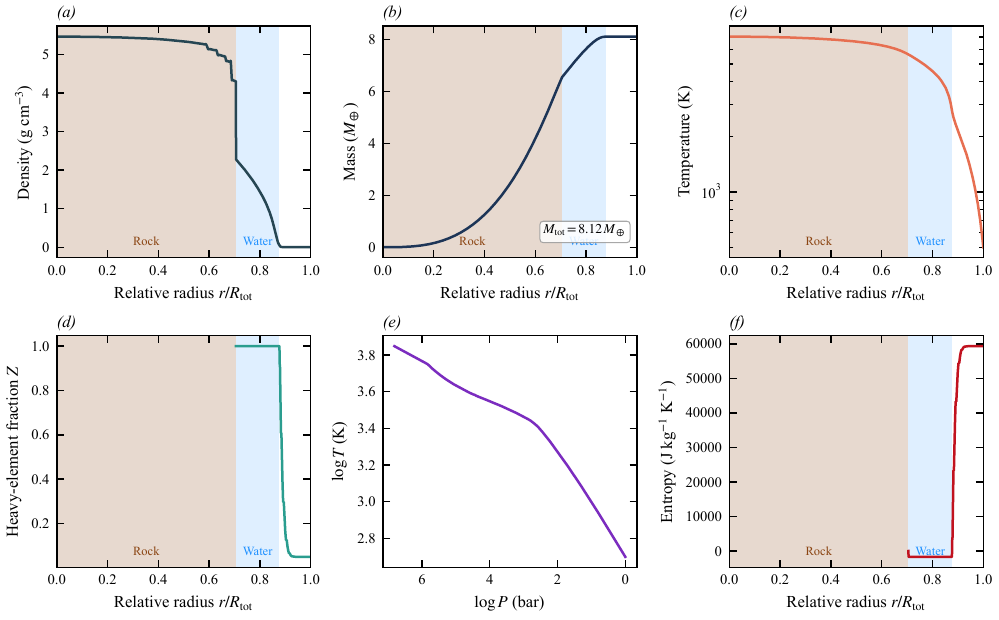}
\caption{Internal structure profile of the GJ\,1214\,b analogue. Panels show, as a function of relative radius, \textit{(a)} density, \textit{(b)} enclosed mass, \textit{(c)} temperature, \textit{(d)} heavy-element fraction $Z$, \textit{(e)} the $P$--$T$ adiabat, and \textit{(f)} specific entropy. The brown shaded region marks the refractory core and the blue shaded region the condensed water mantle.}
\label{fig_gj1214b_app}
\end{figure*}

\begin{figure*}[ht!]
\centering
\includegraphics[width=0.7\textwidth]{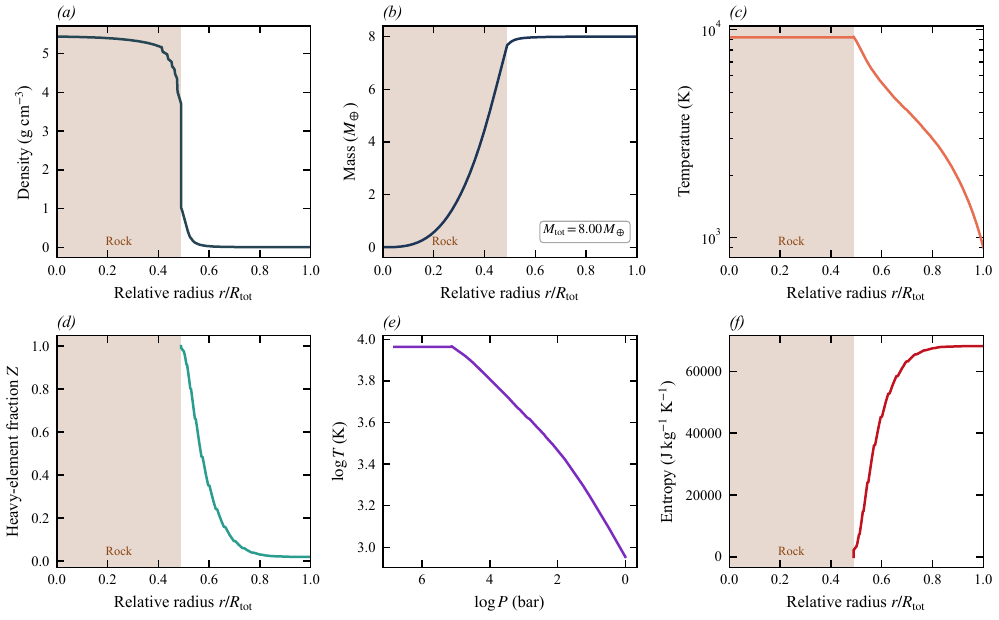}
\caption{Same as Fig.~\ref{fig_gj1214b_app}, but for the Kepler-11 e analogue.}
\label{fig_kepler11e_app}
\end{figure*}

This appendix provides the detailed radial profiles for the two sub-Neptune analogues discussed in Sect.~\ref{Sec:Sub-Neptune Degeneracy}. Fig.~\ref{fig_gj1214b_app} illustrates the dense, volatile-rich architecture of the GJ\,1214\,b analogue, characterised by a massive high-pressure water mantle and a nearly negligible atmospheric envelope. In contrast, Fig.~\ref{fig_kepler11e_app} displays the highly extended architecture of the Kepler-11 e analogue, where a comparable rocky core is instead surrounded by a massive, dilute hydrogen-helium envelope featuring a wide compositional gradient.

The GJ\,1214\,b analogue ($8.12\,M_\oplus$) is dominated by a deep supercritical water mantle extending from the rocky core boundary to a thin gaseous envelope, giving a total radius of $2.72\,R_\oplus$. The Kepler-11 e analogue ($8.00\,M_\oplus$) instead shows a highly extended, dilute gaseous envelope that drives its exceptionally low bulk density and large radius, $4.17\,R_\oplus$.

\subsection{Internal structure profiles of a Jupiter-like planet}
\label{app:jupiter_profile}

\begin{figure*}[ht!]
\centering
\subfloat[Possible internal structure for a Jupiter-like planet]{
    \includegraphics[width=0.66\linewidth]{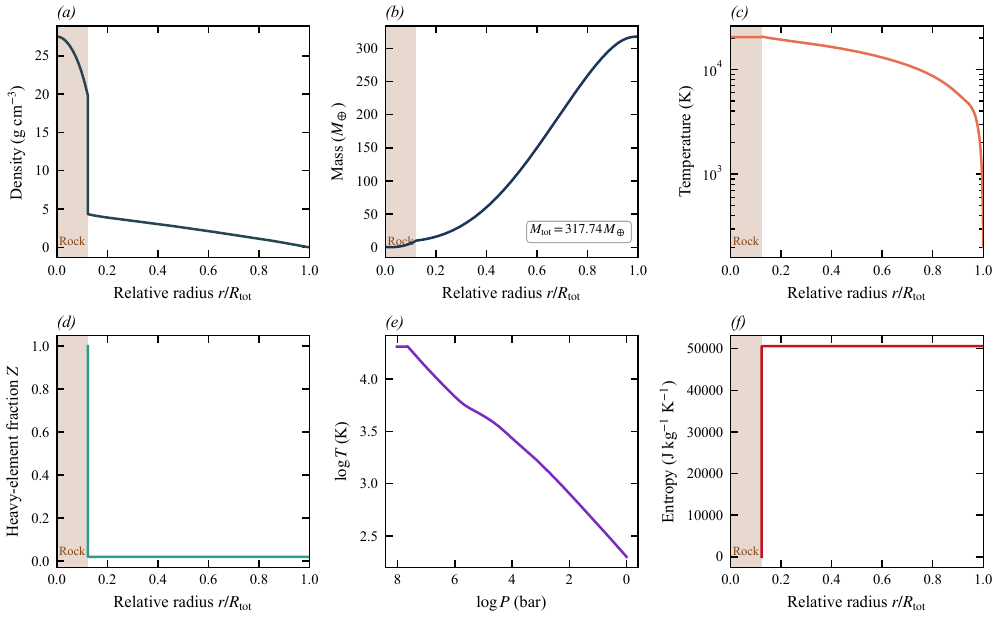}
    \label{fig_amp_curve}
}

\vspace{0.8em}

\subfloat[Planetary $P$--$T$ trajectory overlaid on the compositional phase space]{
    \includegraphics[width=0.66\linewidth]{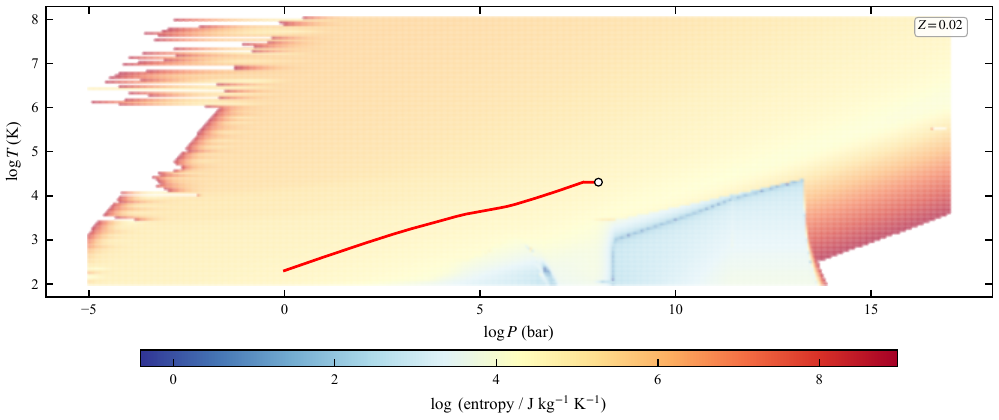}
    \label{fig_amp_bar}
}
\caption{Internal structure and thermodynamic trajectory of a Jupiter-like planet. \textit{(a)} Possible internal structure of a 1 Jupiter-mass planet initiated with a 200\,K temperature at 1\,bar and a continuous adiabat throughout the envelope, giving a radius of 0.97 $R_\mathrm{Jup}$. \textit{(b)} Trajectory in the $P$--$T$ phase space, with specific entropy overlaid on the colour axis.}
\label{fig:jupiter_like}
\end{figure*}

In this appendix, we present the reference structural and thermodynamic profiles for a representative Jupiter-like gas giant. This serves as a baseline comparison for the planetary regimes explored in the main text. Figure~\ref{fig:jupiter_like} details the interior architecture of a 1 Jupiter-mass model, initialised with a boundary temperature of 200\,K at 1\,bar and assuming a continuous adiabatic temperature gradient throughout the gaseous envelope, which yields a derived bulk radius of 0.97 $R_\mathrm{Jup}$. Accompanying the physical structure profile, we trace the corresponding pressure-temperature ($P$--$T$) trajectory across the compositional phase space. This visualisation maps the local entropy variations along the planetary adiabat, highlighting the thermodynamic pathway from the upper atmosphere down to the deep interior.

\subsection{Thermodynamic trajectory in the mixed EOS phase space}
\label{app:phase_space}

\begin{figure*}[ht!]
\centering
\includegraphics[width=0.7\textwidth]{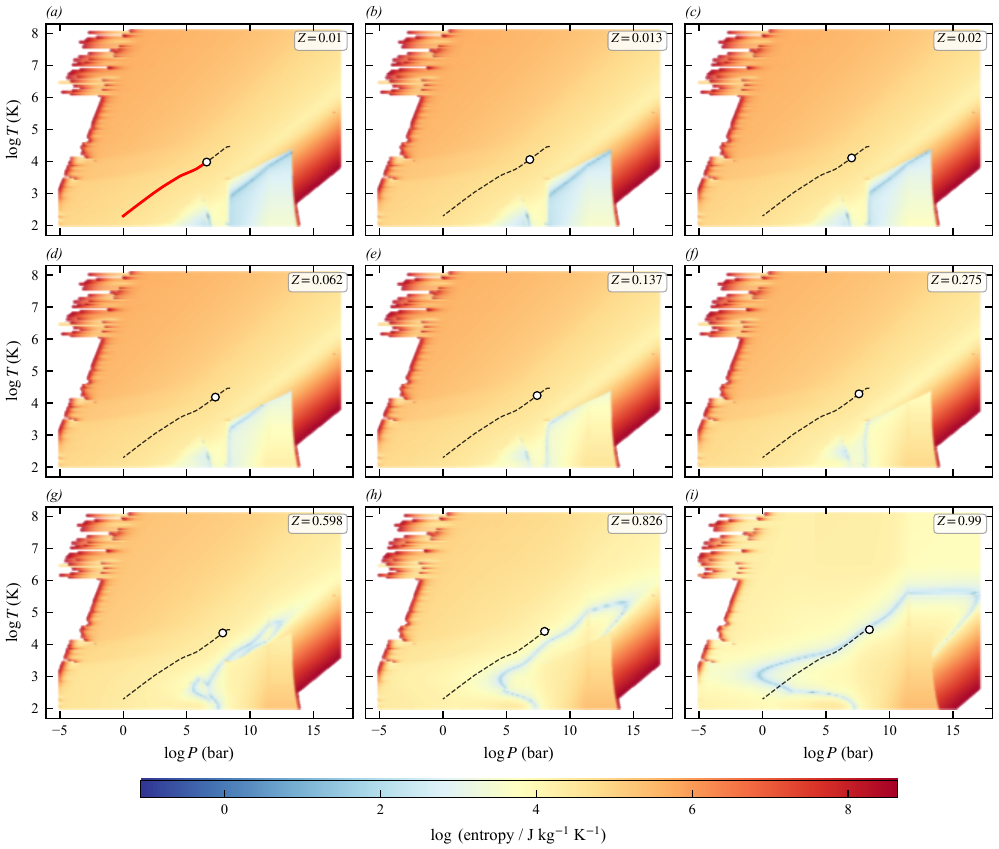}
\caption{Planetary $P$--$T$ trajectory overlaid on the compositional phase space. Each panel corresponds to a different EOS mixing ratio $Z$, given in the panel; the background scatter is the EOS grid for that table, coloured by specific entropy. The red segment indicates the portion of the planet integrated with that table, and the dashed line the full trajectory. The white circle marks the start of each active segment.}
\label{fig_phase_space}
\end{figure*}

Fig.~\ref{fig_phase_space} illustrates the planetary $P$--$T$ trajectory for the $1.0\,M_\mathrm{Jup}$ model with a dilute compositional gradient ($\sigma = 0.05$) discussed in Sect.~\ref{sec:trajectories}. Each panel represents a slice of the 3D mixed EOS space. As the planet becomes enriched in heavy elements deep within the interior, the trajectory passes through tables with progressively higher heavy element mass fractions ($Z$). The smooth transition between these tables, managed by our entropy-matching logic, ensures that the resulting $P$--$T$ adiabat remains physically continuous despite the discrete nature of the numerical staircase.

\subsection{Core temperature gradient sensitivity}
\label{app:core_nabla}

To quantify the structural impact of the assumed thermal profile within
the refractory core,
we ran \texttt{fuzzycore} with the dimensionless core gradient
$\nabla_\mathrm{core} \equiv \mathrm{d}\ln T / \mathrm{d}\ln P$
swept across the range spanned by the three physical limits: isothermal
($\nabla_\mathrm{core} = 0$), radiative-conductive
($\nabla_\mathrm{core} \approx 0.07$, appropriate to electron-degenerate
heat transport in dense matter), and adiabatic ($\nabla_\mathrm{core} = 0.30$).
We tested three gas-giant configurations spanning $0.1$, $0.5$, and
$1.0\,M_\mathrm{Jup}$ with $M_\mathrm{core} = 0.1\,M_\oplus$, plus a
GJ\,1214\,b-like water world ($M_\mathrm{rock} = 6.9$, $M_\mathrm{water} = 1.0\,M_\oplus$).

\begin{figure*}[ht!]
\centering
\includegraphics[width=0.7\textwidth]{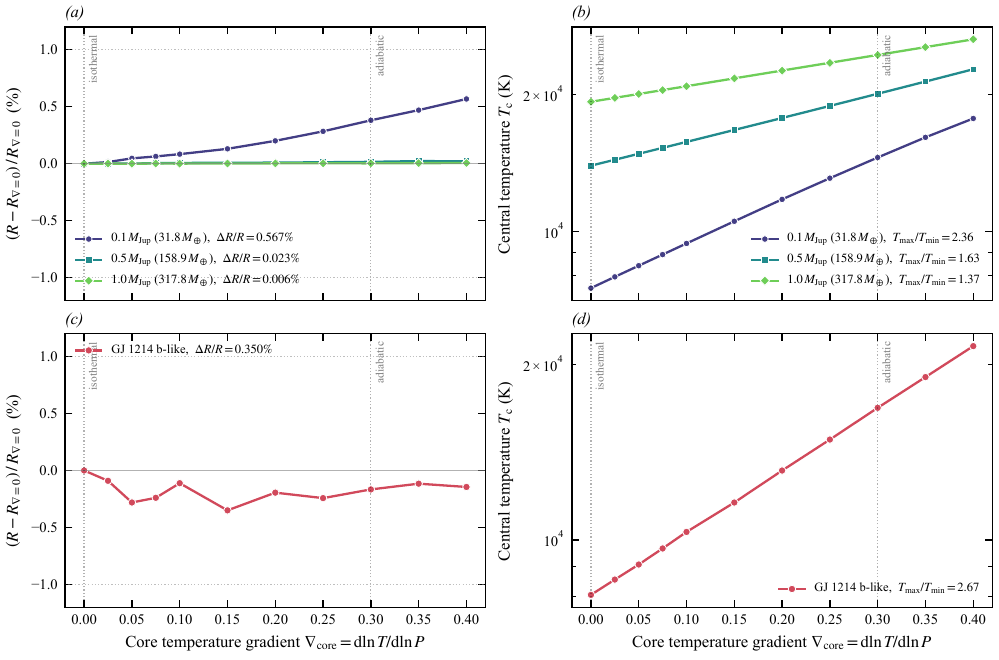}
\caption{Sensitivity of converged planetary structure to the rocky-core
temperature gradient $\nabla_\mathrm{core}$. \textit{(a)} and \textit{(b)}: gas-giant family at $0.1$, $0.5$, and $1.0\,M_\mathrm{Jup}$. \textit{(c)} and \textit{(d)}: GJ\,1214\,b-like water world. Left panels show the relative radius deviation from the isothermal ($\nabla_\mathrm{core} = 0$) baseline; right panels the corresponding central temperature on a log scale. The $\pm 1\%$ dotted reference lines on the radius panels show that the structural bound is respected by a comfortable margin. Dotted vertical lines mark the isothermal and adiabatic limits.}
\label{fig:core_nabla}
\end{figure*}

The results are presented in Fig.~\ref{fig:core_nabla}. Across the full
sweep of $\nabla_\mathrm{core}$, the converged total radius varies by
$0.006\%$ at $1.0\,M_\mathrm{Jup}$, $0.023\%$ at $0.5\,M_\mathrm{Jup}$,
$0.57\%$ at $0.1\,M_\mathrm{Jup}$, and $0.35\%$ in the water world.
Meanwhile, the central temperature varies by factors of $1.4$--$2.4$ in
the gas giants and $2.7$ in the water world. The dramatic underlying
thermodynamic response confirms that the test exercised the core thermal
profile non-trivially; the near-flat radius response reflects that the
condensed iron-silicate equation of state is essentially temperature-independent at megabar pressures
\citep{seager_massradius_2007, zeng_massradius_2016}.

\subsection{Water-mantle temperature gradient sensitivity}
\label{app:mantle_nabla}

The condensed volatile mantle is the only \texttt{fuzzycore} interior
layer whose equation of state is significantly less pressure-dominated
than the refractory core: water at the megabar pressures of sub-Neptune
mantles retains non-negligible thermal expansivity through the Ice~VII,
Ice~X, and superionic transitions \citep{mazevet_ab_2019}. To test
whether this introduces structural sensitivity beyond that quantified
for the core (Appendix~\ref{app:core_nabla}), we extended the integrator
to accept the mantle temperature profile as a controllable input with
three modes: adiabatic (entropy-matched to the envelope, the
historical default); isothermal ($T_\mathrm{mantle} = T_\mathrm{int}$);
and polytropic ($T_\mathrm{mantle}(P) = T_\mathrm{int}\,(P/P_\mathrm{int})^{\nabla_\mathrm{mantle}}$)
swept across $\nabla_\mathrm{mantle} \in [0, 0.30]$.

\begin{figure*}[ht!]
\centering
\includegraphics[width=0.7\textwidth]{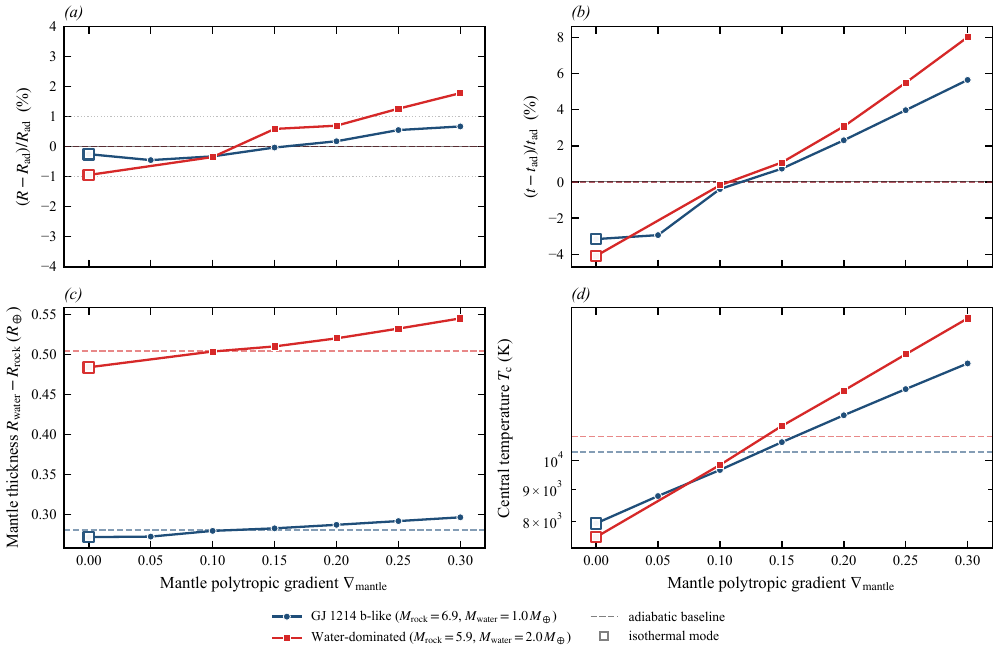}
\caption{Sensitivity of converged sub-Neptune structure to the water-mantle thermal profile. \textit{(a)} Total radius relative to the adiabatic baseline. \textit{(b)} Mantle thickness relative to the same baseline. \textit{(c)} Absolute mantle thickness $R_\mathrm{water} - R_\mathrm{rock}$. \textit{(d)} Central temperature on a log scale. The mantle thickness shows the largest fractional response and the total radius the smallest. The isothermal mode (open square) coincides exactly with polytropic $\nabla_\mathrm{mantle} = 0$, a structural sanity check; dashed horizontal lines mark the adiabatic baseline for each configuration.}
\label{fig:mantle_nabla}
\end{figure*}

For a canonical GJ\,1214\,b-like configuration ($M_\mathrm{rock} = 6.9$,
$M_\mathrm{water} = 1.0\,M_\oplus$), the full sweep produces a total
radius variation of $1.1\%$, a mantle thickness variation of $\sim 9\%$,
and a central temperature contrast of factor $1.8$. For a water-dominated
configuration ($M_\mathrm{rock} = 5.9$, $M_\mathrm{water} = 2.0\,M_\oplus$),
the radius response rises to $\sim 2.7\%$, the mantle thickness response
to $\sim 12\%$, and the central temperature contrast to factor $\sim 2$
(Fig.~\ref{fig:mantle_nabla}). The polytropic mode reproduces the
isothermal limit exactly at $\nabla_\mathrm{mantle} = 0$ (a built-in
sanity check confirmed to six decimal places of agreement), and
converges towards the entropy-matched adiabatic baseline near
$\nabla_\mathrm{mantle} \approx 0.10$--$0.15$, consistent with the local
adiabatic index of the \citet{mazevet_ab_2019} water EOS at the relevant interface
pressures.
The mantle thickness is therefore the most responsive structural piece
in the layered code, but its $5$--$12\%$ variation is diluted to the
$1$--$3\%$ level in the total radius once the pressure-dominated rocky
core and the H/He envelope are accounted for.

\subsection{Solver failure topology}
\label{app:failure_topology}

To demonstrate the numerical limits and convergence robustness of the \texttt{fuzzycore} integrator, we present the failure topology of the sub-Neptune degeneracy atlas (Sect.~\ref{sec:atlas}). Out of $2700$ sampled configurations across the $(M_\mathrm{water}, M_\mathrm{env}, \sigma)$ parameter space for an $8\,M_\oplus$ planet, $80\%$ converged successfully. 

\begin{figure*}[ht!]
\centering
\includegraphics[width=0.7\textwidth]{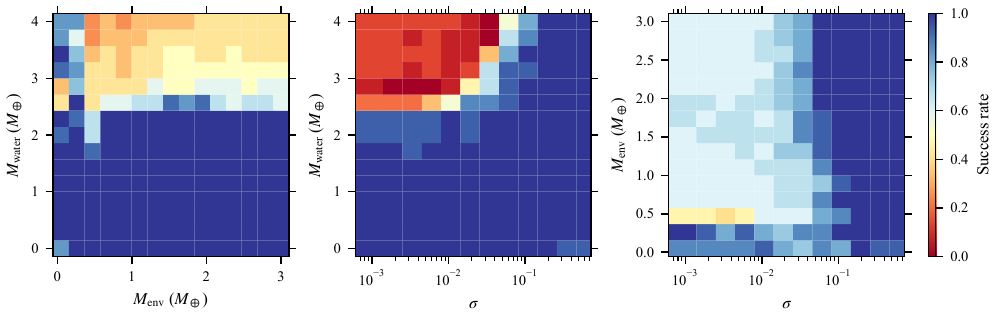}
\caption{Solver failure topology for the $8\,M_\oplus$ water-world atlas, showing the success rate marginalised over the three grid dimensions: $(M_\mathrm{water}, M_\mathrm{env})$ (left), $(M_\mathrm{water}, \sigma)$ (middle), and $(M_\mathrm{env}, \sigma)$ (right). Blue cells denote $100\%$ convergence, red cells complete solver failure. Failures are highly localised to the unphysical extremes of ultra-thin envelopes overlying massive condensed mantles.}
\label{fig:failure_topology}
\end{figure*}

Fig.~\ref{fig:failure_topology} maps the success rate marginalised over the grid dimensions. Failures are not random; they are strictly clustered in the unphysical extremes of the parameter space. Specifically, the solver struggles at the combination of vanishingly thin envelopes ($M_\mathrm{env} < 0.1\,M_\oplus$) overlying massive condensed mantles ($M_\mathrm{water} \to 4.0\,M_\oplus$). In this extreme regime, the pressure scale height of the envelope becomes so diminutive that resolving the compositional gradient triggers chaotic temperature divergence in the EOS interpolator. These regions are physically degenerate with stripped cores and are typically handled via distinct atmospheric boundary assumptions rather than continuous fluid integrations. The high success rate across the bulk of the parameter space validates the stability of the nested boundary-shooting method in resolving generic compositionally graded interiors.

\end{appendix}

\end{document}